\documentclass[11pt]{combine}
\usepackage{amsmath,amsfonts,amssymb,amsthm,epsfig, graphicx, hyperref}
\usepackage[dvipsnames,table,dvipsnames*, svgnames*, hyperref]{xcolor}
\usepackage{natbib,amsmath,amssymb,amsthm,graphicx,setspace,paralist,booktabs,rotating,subcaption,float,color}
\usepackage{multirow}
\usepackage{fancyhdr}
\usepackage{bibunits}
\usepackage[small]{titlesec}

\newtheorem{proposition}{Proposition}
\newtheorem{assumption}{Assumption}

{
	\theoremstyle{definition}

}

\usepackage{booktabs,caption}
\usepackage[flushleft]{threeparttable}

\makeatletter			
	\newcommand*{\indep}{%
		\mathbin{%
			\mathpalette{\@indep}{}%
		}%
	}
	\newcommand*{\nindep}{%
		\mathbin{%                   % The final symbol is a binary math operator
			\mathpalette{\@indep}{\not}% \mathpalette helps for the adaptation
		}%
	}
	\newcommand*{\@indep}[2]{%
		\sbox0{$#1\perp\m@th$}%        box 0 contains \perp symbol
		\sbox2{$#1=$}%                 box 2 for the height of =
		\sbox4{$#1\vcenter{}$}%        box 4 for the height of the math axis
		\rlap{\copy0}%                 first \perp
		\dimen@=\dimexpr\ht2-\ht4-.2pt\relax
		\kern\dimen@
		{#2}%
		\kern\dimen@
		\copy0 %                       second \perp
	} 
	\makeatother	
	
\begin{document}

%\begin{titlepage}

\title{\scshape Improving efficiency of inference in clinical trials with external control data}
\author{Xinyu Li$^1$, Wang Miao$^1$, Fang Lu$^2$ and Xiao-Hua Zhou$^3$ \\
School of Mathematical Sciences \& Center for Statistical Science, Peking University$^1$\\
Xiyuan Hospital, China Academy of 
Chinese Medical Sciences$^2$\\
Department of Biostatistics \& Beijing International Center for Mathematical Research, \\Peking University$^3$\\
Beijing, China 100871}
\date{}
\maketitle
\thispagestyle{empty}

%  put the summary for your paper here

	\begin{abstract}
Suppose we are interested in the effect of a treatment in a clinical trial. The efficiency of inference may be limited due to small sample size. However, external control data are often
available from  historical   studies. Motivated by an application to Helicobacter pylori infection, we show how to borrow strength from such data to improve efficiency of inference  in the clinical trial.  Under an exchangeability assumption about the potential outcome mean, we show that the semiparametric efficiency bound for estimating the 
average treatment effect can be reduced by   incorporating  both  the clinical trial data and  external controls.  
We then derive a doubly robust and locally  efficient estimator.  The improvement in efficiency  is prominent especially when the external control dataset has a large sample size and  small variability. 
Our method  allows for a relaxed overlap assumption, and we illustrate with the case where the clinical trial only  contains a treated group.   We also develop  doubly robust and locally efficient approaches that extrapolate the causal effect in the clinical trial to the external  population and the overall population. Our results also offer a meaningful implication for trial design and data collection.
We evaluate the finite-sample performance of the proposed estimators via simulation. 
In the Helicobacter pylori infection application, our approach shows  that the combination treatment has potential efficacy advantages over the triple therapy.
	\bigskip
	
	\noindent\emph{Key words}: Data combination; Double robustness; Efficiency; External control.
\end{abstract}

\section{Introduction}\label{intro.sec}
%%%%%%%%%%%%%%

Although  randomized clinical trials are the gold standard for evaluation of a treatment effect,  the efficiency of inference may be limited due to small sample size. However, external datasets containing control arm data are increasingly available from historical clinical databases. 
As a motivating application, we consider   a clinical study on Helicobacter pylori (H.pylori) infection, which is a leading world-wide infectious disease. 
A randomized clinical trial is conducted at the traditional Chinese medicine (TCM) hospital, with the aim of examining  whether additional use of TCM can lead to better efficacy than only the triple therapy (clarithromycin, amoxicillin and omeprazole), a standard of care for H.pylori infection.
But the sample size of this clinical trial is small.  In parallel, external control data are available from a single-arm study  at the Western-style hospital where all patients receive the triple therapy; see section 5 for more details about the application. These external control data have the potential to improve the power of statistical inference about the treatment effect.

Borrowing strength from external control data to  improve inference  has long been a demand in the design and analysis of clinical trials since  \cite{pocock1976combination}, termed as {\it external control},  {\it historical control} or {\it historical borrowing} in the literature; it is also mentioned in clinical  guidance documents \citep[e.g.,][]{fda2017}. For recent practice of using external control data in  clinical and pharmaceutical studies, see \cite{viele2014use,van2018including, schmidli2019beyond}.  
As a straightforward approach, synthesizing clinical trial  and external control data as if they were from the same population may lead to bias, due to heterogeneity across study populations. 
Several other methods, such as common meta-analytic, Bayesian and frequentist methods \citep[e.g.,][]{viele2014use,weber2018use,zhang2019bayesian} are also shown by \cite{li2019target} to fail to adequately address population heterogeneity.

In recent years, there has been an increasing amount of literature on causal inference methods that integrate  information from multiple data sources, particularly when both randomized trials and observational studies are available. Most of these studies focus on {\it extrapolating} the causal inference from the randomized trials, such as drawing causal inference over a target population \citep{stuart2011use,hartman2015sample, zhang2016new,rudolph2017robust,buchanan2018generalizing,rosenman2018propensity,chen2019doubly, dahabreh2019generalizing,dahabreh2020toward,dahabreh2020extending,li2021generalizing}, or validating observational methods which may suffer from confounding bias \citep{kallus2018removing,lodi2019effect,athey2020combining,yang2020doubly}.
Much of the previous research in this field is developed in nested designs; for example, \cite{dahabreh2019generalizing} investigate the case where trials are nested in cohort studies that collect baseline data from all eligible individuals, and \cite{lu2019causal} consider a comprehensive cohort study; yet recently non-nested designs have received growing attention where observations in different datasets are sampled separately.  \cite{pearl2011transportability,pearl2014external} also consider the data integration using directed acyclic graphs. Different terms are often used in literature  for similar issues, e.g., 
{\it generalizability}, {\it external validity},
{\it transportability}, and  {\it data fusion}, see \cite{colnet2020causal} for a systematic review.

In this article, we show how to improve efficiency of inference in the clinical trial by {\it borrowing} strength from external control data.
Analogous to \cite{dahabreh2019generalizing,dahabreh2020extending}, we maintain a  mean exchangeability assumption, which reveals that  the mean efficacy of the placebo or a standard of care is the same across  studies within  the same baseline level.
Under the mean exchangeability,  we derive  the semiparametric efficiency bound for estimating the average treatment effect in the clinical trial when external controls are available. 
It is lower than the bound obtained without  external controls. 
For estimation, we develop a semiparametric estimator, which is doubly robust and locally efficient. 
The efficiency gain over the  trial-based doubly robust estimator is significant especially when the external control dataset has a large sample size and small noise and when there is a sufficient overlap between the two datasets regarding covariates. Our method  allows for a relaxed overlap assumption, which is weaker than that required by the trial-based estimator. 
We also develop doubly robust estimators that generalize the causal inference from the trial population  to the  external population and to the overall population of the study. 
The  two estimators are more efficient than the estimator  proposed  by \cite{rudolph2017robust,dahabreh2020extending} and the one  proposed  by \cite{dahabreh2019generalizing} respectively, as we make full use of the outcome observations contained  in the external dataset.  We illustrate with  simulations  and an  application to the H.pylori study, which is of a non-nested design. In the application, our estimators result in  smaller variances and  show  that the combination treatment has potential efficacy advantages over the triple therapy.

\section{Inference about the treatment effect in the clinical trial}
\label{cepd}

\subsection{Study design and data structure}

In general, the study designs for multiple observational datasets fall into two categories: 
(i) nested trial designs, %in which the trial sample is     nested in a large sample that is selected in advance from the target population;
(ii) non-nested trial designs, %in which observations in different datasets are sampled separately, 
see \cite{dahabreh2020extending,dahabreh2021study,colnet2020causal}. %The sampling probabilities are generally known in nested trial designs, but unknown in non-nested trial designs. 
Our motivating H.pylori application concerns a non-nested composite dataset design. 
To understand the sampling model in the composite dataset design, we consider the following setup.  Suppose that there exists an underlying population, i.e., the study-eligible patient population of H.pylori infection to whom research studies would be applicable. The patient preference defines two different sub-populations,  namely         the study-eligible patient population at the TCM hospital and that at the Western-style hospital. For simplicity,  we refer to them as the trial population and external population, respectively. 
Then the clinical trial observations and external data observations can be viewed as simple random samples from the corresponding sub-populations with unknown sampling probabilities.    
Therefore, each sub-population is identified by its sampled observations, but the underlying patient population is not due to the unknown sampling probabilities. 
The overall population  corresponding to all observed samples  is a mixture of the two sub-populations as we discuss later, yet is different from the underlying population in general. See the Supporting Information for  detailed  descriptions of the design and discussions about the relationship to other designs.
Note that our proposed methods also apply to other designs as long as the data structure and assumptions described below are satisfied; we will return to this issue in the discussion.

Now we formally describe the observed data structure in our setting.  We let $Y$ denote the outcome of interest, $T$ the indicator for treatment assignment with $T=1$ for the treated group and $T=0$ the control group, $D$ the indicator for data source with $D=1$ for the clinical trial and $D=0$ for external control, and $X$ a set of pre-treatment covariates.  By the design, the observed dataset $\mathcal{O}$ consists of $n$ independent and identically distributed samples of $(Y,X,T,D )$ from the overall population; it is divided into two separate subsets, the clinical trial dataset $\mathcal{O}_1$ containing $n_1$ observations  of $(Y,X,T,D=1)$, and the external control dataset $\mathcal{O}_2$ containing $n_2=n-n_1$ observations of $(Y,X,T=0,D=0)$. Table 1 illustrates the observed data structure. The limitation of $n_1/n$ as $n \rightarrow \infty$ approaches a positive constant $q={\rm pr}(D=1)$.  
We maintain the  stable unit treatment value assumption that no interference between units  and no hidden variations of treatments occur. We let  $Y_t$ denote  the potential outcome that would be observed had $T$ been set to $t$ for $t\in\{0,1\}$.  We make the consistency assumption  that the observed outcome is a realization of the potential outcome under the exposure actually received.
\begin{assumption}\label{assump0}
\begin{itemize}
\item[(i)] $Y=TY_1+(1-T)Y_0$ for observations in $\mathcal{O}_1$;
\item[(ii)] $D=0 \Longrightarrow T=0$ and thus $Y=Y_0$ for observations in $\mathcal{O}_2$.
\end{itemize}
\end{assumption}
In this section, the causal parameter of interest is $\tau=E(Y_1-Y_0\mid D=1)$, the average treatment effect  in the trial population.

\begin{table}
 \centering
 \def\~{\hphantom{0}}
  \begin{minipage}{175mm}
  \caption{Observed data structure.  ``$\checkmark$" and ``$?$" indicate observed and unobserved, respectively}
  	\label{tab1}
 \begin{tabular}{cccccccc} 
               		\toprule 
               &	&Dataset	&Treatment  & Covariates & Potential \hspace{-0.4em} & \hspace{-1.6em} Outcome     &\hspace{-1em} Observed Outcome \\
               &	&	$D$ & $T$ & $X$ & $Y_1$ & \hspace{1.2em} $Y_0$ \hspace{1.2em} & $Y$ \\ \toprule 
              &  $1$	&	1 & 1 & $\checkmark $ & $ \checkmark $ & ? & $ \checkmark $ \\ 
        $\mathcal{O}_1$    &    $\vdots$	&	$\vdots$ & $\vdots$ & $\vdots$ & $\vdots$ & $\vdots$ & $\vdots$\\ 
            &   $n_1$	&	1 & 0 & $ \checkmark $ & ? & $ \checkmark $  & $ \checkmark $ \\ \toprule 
            &    $n_1+1$	&	0 & 0 & $ \checkmark $ & ? & $ \checkmark $ &$ \checkmark $\\ 
        $\mathcal{O}_2$   &     $\vdots$	&	$\vdots$ & $\vdots$ & $\vdots$ & $\vdots$ & $\vdots$ & $\vdots$\\ 
            &   $n$	&	0 & 0 &$ \checkmark $& ? & $ \checkmark $& $ \checkmark $\\ 
               		\toprule 
               	\end{tabular}
 \end{minipage}
\end{table}

\vspace{-1em}

\subsection{Inference solely based  on the clinical trial data}

Identification of the treatment effect $\tau$ is guaranteed based on the  trial data under assumption \ref{assump0} and the following strong ignorability assumption \citep{rosenbaum1983central}, which is commonly required in observational studies and can be met in randomized trials.

\begin{assumption}[Strong ignorability]\label{assump1}
\begin{itemize}
\item[(i)] Ignorability: $(Y_1,Y_0)\indep T \mid X,D=1$;
\item[(ii)] Overlap: $0\hspace{-0.1em} <\hspace{-0.1em} \mbox{pr}(T=1 \mid X=x,D=1) \hspace{-0.1em} < \hspace{-0.1em}1$ for all $x$ such that $ \mbox{pr}(X=x\mid D=1)>0$.
\end{itemize}
\end{assumption}

We let $m_t(X) =  E(Y \mid X,D=1,T=t)$ for $t\in \{0,1\}$ and  $p(X) = \mbox{pr}(T=1 \mid X,D=1)$ denote the outcome model and the treatment propensity score model, respectively. 
The residual of $Y$ by subtracting  $m_t(X)$ is $R_t = Y-m_t(X)$, and the  difference between $m_t(X)$ is  $\Delta(X)=m_1(X)-m_0(X)$.
Under assumptions \ref{assump0}-\ref{assump1}, the asymptotic variance of any regular and asymptotic linear estimator of $\tau$ based solely on the trial data    can be no smaller than the  efficiency bound $\tilde{{B}}_\tau=E(\tilde{\mathrm{IF}}_\tau^2)$   \cite[see][]{robins1994estimation,hahn1998role,van2003unified}, where
\begin{equation*} 
\begin{aligned}
\tilde{\mathrm{IF}}_\tau=\frac{D}{q}&\left\{\Delta(X)-\tau+\dfrac{T}{p(X)} R_1-\dfrac{1-T}{1-p(X)}R_0\right\},
\end{aligned}
\end{equation*}
is the efficient influence function for $\tau$ when only the trial data are used for estimation.

We let  $m_t(X;\beta_t)$ and $p(X; \phi)$   denote  parametric working  models for $m_t(X)$ and $p(X)$, respectively.  
A doubly robust estimator of $\tau$  can be  obtained by first fitting $m_t(X;\tilde \beta_t)$ and $p(X; \tilde \phi)$ with the clinical trial data and then using $\tilde{\mathrm{IF}}_\tau=0$ as an estimating equation:
\begin{equation*}
\begin{aligned}
\tilde {\tau}_{\mathrm{dr}}=\hat{E}\left[\left\{\tilde{\Delta}(X) 
+ \dfrac{T}{p(X;\tilde{\phi})}\tilde{R}_1 -  \dfrac{1-T}{1-p(X;\tilde{\phi})}  \tilde{R}_0 \right\}\mid  D=1 \right],
\end{aligned}
\end{equation*}
where $\hat{E}$ is the empirical mean operator, $\tilde{\Delta}(X)=m_1(X;\tilde{\beta}_1)-m_0(X;\tilde{\beta}_0)$,
and $\tilde{R}_t=Y-m_t(X;\tilde{\beta}_t)$ for $t\in\{0,1\}$. 
This estimator only rests on the observed data from the clinical trial, and thus we refer to $\tilde \tau_{\rm dr}$ as  the trial-based doubly robust estimator.

Under assumptions \ref{assump0}-\ref{assump1} and  regularity conditions  \cite[theorems 2.6 and 3.4]{newey1994large},   $\tilde{\tau}_{\mathrm{dr}}$ is doubly robust in the sense that it is consistent if either set of the working  models are correctly specified. In recent years, such estimators have grown in popularity for inference about missing data and treatment effects \citep{scharfstein1999adjusting}. 
Moreover, if both  working models are correct, the asymptotic variance of the estimator $\tilde {\tau}_{\mathrm{dr}}$ attains the efficiency bound $\tilde{B}_\tau$  \citep{bang2005doubly,cao2009improving}.
However,  in the presence of external control data, the efficiency bound for estimating $\tau$  can be lower than $\tilde{B}_\tau$, and thus  $\tilde{\tau}_{\mathrm{dr}}$  is no longer efficient because it  does not  take external controls into account. 
We aim to  characterize the semiparametric efficiency bound of  $\tau$ when  external control data are available. 

\subsection{ Improving efficiency   with external control data}\label{esti}

In order to incorporate external control data, we maintain the following assumption.
\begin{assumption}[Mean exchangeability for $Y_0$]
\label{assump2}
$E(Y_0\mid X,D=0)=E(Y_0 \mid X,D=1)$.
\end{assumption}

Assumption \ref{assump2} reveals that given  pre-treatment covariates, the mean efficacy of the placebo or a standard of care remains the same in the clinical trial and external control studies. 
It implicitly excludes average  engagement effects caused by participation in a particular study rather than by treatment, for example, the so-called Hawthorne effects. 
Such an assumption is adopted in  many practical applications   \citep[e.g.,][]{rudolph2017robust,dahabreh2019generalizing,dahabreh2020extending}. 
In our H.pylori application, it is plausible because the control treatment is a standard of care in clinical  settings regardless of the type of hospital and participation of the study.
Relating this to the transport formula  developed by \cite{pearl2014external}, this assumption means that the selection  node ($D$) does not point into the  outcome node ($Y$).
When assumptions  \ref{assump0} and \ref{assump1} hold,  this assumption implies   $E(Y\mid X,D=0,T=0)=E(Y \mid X,D=1,T=0)$,
which is a testable implication of the mean exchangeability.
To assess assumption \ref{assump2}, 
we can fit a parametric outcome model for $E(Y\mid X,D,T=0)$ to test interactive effects between $D$ and $X$, or see \cite{luedtke2019omnibus} for a nonparametric omnibus test. 
The mean exchangeability  is  weaker than  the distribution exchangeability $Y_0\indep D \ |\ X$,  considered by \cite{stuart2011use,hartman2015sample,buchanan2018generalizing,breskin2019using, lu2019causal,li2019target,yang2020doubly}.

We aim to calculate the semiparametric efficiency bound for estimating $\tau$ in the nonparametric model,  where  the observed data distribution is only restricted by assumptions \ref{assump0}-\ref{assump2}.
We let $\pi(X) = \mbox{pr}(D=1 \mid X)$ denote the selection propensity score, which quantifies the difference between  participants in the trial  and external data, and  let $r(X)= \mbox{var}(Y_0 \mid X,D=1)/\mbox{var}(Y_0 \mid X,D=0)$, which measures the relative variability of $Y_0$ 
between the trial and external data. 
Under assumptions \ref{assump0}-\ref{assump1}, the variance ratio $r(X)$ equals $\mbox{var}(Y \mid X,D=1,T=0)/\mbox{var}(Y \mid X,D=0)$.
In certain cases, $r(X)$ can be known;  for a binary outcome,   $r(X)=1$ under assumption \ref{assump2}. 
For the ease of notation,  we let
\begin{equation*}
W(X,D,T)=\dfrac{D(1-T)\pi(X) +(1-D) \pi(X)r(X)}{\pi(X)\{1-p(X)\}+\{1-\pi(X)\}{r(X)}}.
\end{equation*}

\begin{proposition}\label{efficiency}
Under assumptions \ref{assump0}--\ref{assump2}, the efficient influence function for $\tau$
is
\begin{eqnarray*} 
\mathrm{IF}_\tau&=&\dfrac{1}{q}\left[D\{\Delta(X)-\tau\}
+\dfrac{DT}{p(X)} R_1-W(X,D,T)R_0\right],
\end{eqnarray*}
and the semiparametric efficiency bound is ${B}_\tau=E(\mathrm{IF}_\tau^2)$.
\end{proposition}

We compare  $B_\tau$ to $\tilde{B}_\tau$ to show how 
external control data can improve efficiency of inference about the treatment effect in the trial population. 
Let $V_d(X)=\mbox{var}(Y_0\mid X,D=d)$  and in the Supporting Information we show that
\begin{equation}\label{vd}
\tilde{B}_\tau - B_\tau 
=  E\bigg[   \bigg\{ \frac{1}{1-p(X)} - \frac{1}{1-p(X)+\frac{{\rm pr}(X\mid D=0)}{{\rm pr}(X\mid D=1)}\frac{1-q}{q} r(X)}  \bigg\} 
\frac{V_1(X)}{q} \mid D=1 \bigg], 
\end{equation} 
which is  positive  as long as  there exist a non-zero measure set of overlapped $X$ for which ${\rm pr}(X\mid D=1){\rm pr}(X\mid D=0)>0$.

 From equation \eqref{vd}, given a trial dataset drawn from a fixed trial population,
 the efficiency gain brought by the external dataset is determined by three key factors:
 the degree of overlap between covariate distributions in the  trial and external populations, the proportion of external samples in the overall dataset and the variance ratio, captured by 
  ${\rm pr}(X\mid D=0)/{\rm pr}(X\mid D=1)$, $q$ and $r(X)$, respectively.
  For an ideal case where $p(X)$ and conditional variances $V_d(X)$ are constants, in the Supporting Information we show that the optimal ${\rm pr}(X\mid D=0)$ which maximizes the efficiency gain is equal to ${\rm pr}(X\mid D=1)$.  
Besides,   a  smaller $q$ and  a larger $r(X)$ lead to a bigger gap between the two efficiency bounds. Therefore, it is advantageous to collect an external dataset with a large sample size and small noise from an external population that has a sufficient overlap
regarding covariates with the trial population.
Moreover, the efficiency improvement also relates to the treatment assignment mechanism captured by $p(X)$; if the trial dataset contains very few  control units, the efficiency  gain due to external control data is  prominent. These results offer a meaningful implication for trial design and data collection. In the Supporting Information, we illustrate this with an ideal example in which  $B_\tau/\tilde{B}_\tau$ is derived and  more simulations.

\subsection{A novel doubly robust estimator  incorporating external control data}\label{comp}

The efficient influence function $\mathrm{IF}_\tau$ described in proposition 1 motivates an estimator that achieves the semiparamtric efficiency bound  by using $\mathrm{IF}_\tau$ as an estimating equation.
In addition to  the models $m_t(X;\beta_t)$ and   $p(X; \phi)$
used in the trial-based doubly robust estimator,  
we also specify parametric   working models $\pi(X; \alpha)$ for the selection propensity score $\pi(X) $ and $r(X;\eta)$ for the variance ratio  $r(X)$.  
Under assumptions \ref{assump0}--\ref{assump2}, $m_0(X)=E(Y\mid X,T=0)$, and thus we  fit the outcome model $m_0(X;\beta_0)$ with all control units available from both the trial and external data.
We let $(\hat{\beta}_t,\hat{\phi},\hat{\alpha},\hat{\eta})$ denote a $n^{1/2}$-consistent estimator of the nuisance parameters.  
The efficient influence function motivates the following estimator of $\tau$,
 \begin{equation*}
\hat{\tau}_{\mathrm{dr}}=\frac{1}{\hat{q}}\hat{E}\bigg{[}
 D\widehat\Delta(X)
+ \left. \dfrac{DT}{p(X;\hat{\phi})}\widehat R_1 - \widehat W(X,D,T)\widehat R_0\right],
\end{equation*}
where $\hat q=n_1/n$, $ \widehat\Delta(X)=m_1(X;\hat{\beta}_1)-m_0(X;\hat{\beta}_0)$ and $\widehat R_t=Y-m_t(X;\hat{\beta}_t)$ for $t\in\{0,1\}$ and 
\[  \widehat W(X,D,T)  = \dfrac{D(1-T) + (1-D)r(X;\hat{\eta})}{{\pi}(X;\hat{\alpha})\{1-p(X;\hat{\phi})\}+\{1-{\pi}(X;\hat{\alpha})\}r(X;\hat{\eta})} {\pi}(X;\hat{\alpha}).\]

Under standard regularity conditions  \citep[theorem 25.54]{van2000asymptotic}, the proposed estimator $\hat{\tau}_{\mathrm{dr}}$  is locally efficient, in the sense that its asymptotic variance attains the semiparametric efficiency  bound ${B}_\tau$ when all working models are correctly specified.
Moreover, $\hat{\tau}_{\mathrm{dr}}$ is also doubly robust, just as many locally efficient estimators are naturally robust for estimating parameters that arise in causal inference  \citep[see e.g.,][]{benkeser2017doubly}.

\begin{proposition}\label{p1}
Under assumptions~\ref{assump0}--\ref{assump2} and  regularity conditions described in 
theorems 2.6 and 3.4 of  \citet{newey1994large}, the
 estimator $\hat{\tau}_{\mathrm{dr}}$  is  consistent and asymptotically normal if either 
(i) the outcome models  $m_t(X;\beta_t)$ for $t \in \{0,1\}$ are correct,  or 
(ii) the propensity score models $\pi(X; \alpha)$ and $p(X; \phi)$ are correct.
\end{proposition}
The proposed estimator $\hat{\tau}_{\mathrm{dr}}$ incorporates all data available from both the trial dataset and 
the external controls, and thus we call it the full-data doubly robust estimator to distinguish from the trial-based  one $\tilde \tau_{\rm dr}$. 
Although we focus on parametric working models, the full-data doubly  robust estimator can  accommodate  flexible working models and    leads to  a valid  $n^{1/2}$ inference as long as estimators of nuisance parameters have a convergence rate of at least $n^{1/4}$ \citep{newey1990semiparametric, robins2017higher}. Such a rate is achievable by a bunch of machine learning methods, such as  random forests \citep{wager2018estimation}, neural networks \citep{chen1999improved}, and the highly adaptive lasso \citep{benkeser2016highly}. 

%Besides, the efficient influence function  also suggests a one-step bias correction, by which the efficient estimator can be easily obtained based on an initial consistent estimator, see \citet{bickel1997efficient} and \citet{benkeser2017doubly}.  

Compared to the trial-based estimator  $\tilde \tau_{\rm dr}$, the full-data estimator  $\hat \tau_{\rm dr}$ entails two extra working models $\pi(X;\alpha)$ and  $r(X; \eta)$. 
Consistency of  $\hat \tau_{\rm dr}$ rests on  the selection propensity score model $\pi(X;\alpha)$, which  is essential  for incorporating the external control data.
However, consistency  of  $\hat \tau_{\rm dr}$ does not rest on the variance ratio model $r(X; \eta)$, although the efficiency does. 
The variance ratio $r(X)$ strikes a  balance between the control units in the trial and external data by assigning  more weights to those with smaller noise.  
A larger $r(X)$ attaches more importance to  the external data.
If  $r(X)$ is replaced by zero and $\beta_0$  is estimated with only the trial data, then  external control data are ignored and $\hat \tau_{\rm dr}$ reduces to the trial-based  estimator $\tilde {\tau}_{\mathrm{dr}}$.
In certain cases,  the variance ratio is known and $\hat{\tau}_{\mathrm{dr}}$  reduces to a simplified  form.   
For instance, when the outcome is binary or $Y_0 \indep D \mid  X$ holds, we have  $r(X)=1$.

Although we require  that $p(X)<1$ in assumption  \ref{assump1},  it can be   relaxed to $p(X) \pi(X)<1$, which is sufficient for the denominator  $\pi(X)\{1-p(X)\}+\{1-\pi(X)\}{r(X)}$ in $\mathrm{IF}_\tau$  to be positive.
It indicates that for the treated units,  similar control units only need to exist  in either the clinical trial or the external data. In the Supporting Information, we discuss this issue in  detail  and consider  an extreme  case where the trial dataset only contains a treated group. In this case, we let $p(X)=1$, then the full-data estimator $\hat{\tau}_{\mathrm{dr}}$  reduces to an estimator which is doubly robust against misspecification of either  $m_0(X;\beta_0)$ or $\pi(X;\alpha)$.

In addition to  point estimates,   standard errors and confidence intervals can be obtained  based on normal approximations under certain regularity conditions, which follows from the general theory for estimation equations \citep{van2000asymptotic}. Alternatively, bootstrap methods could also be implemented to obtain the variance  estimate.

\section{Extrapolating inference to other populations of interest}\label{overall}

In section 2, we have focused on the treatment effect in a clinical trial.
However, the treatment effects in the   external and the overall population may differ and thus extrapolation of  inference is of interest. By the study design, the overall population can be considered as a mixture of the trial and external populations, with the mixing probability $q$. This mixed population characterizes patients in both types of hospitals and, although generally not  equivalent to the underlying population, is relatively more representative than a single sub-population.
In this section, we aim to  develop locally efficient and doubly robust estimators of  $\xi=E(Y_1-Y_0\mid D=0)$ and $\psi=E(Y_1-Y_0)$.

However, $\xi$ and $\psi$ are not identifiable without further assumptions, because the potential outcome $Y_1$  is not observed in the external control data. To aid in identification, we assume an additional  mean exchangeability for  $Y_1$
and an  additional overlap condition.

\begin{assumption}[Mean exchangeability for $Y_1$]
\label{assump3}
$E(Y_1\mid D=0,X)=E(Y_1\mid D=1,X)$.
\end{assumption}

\begin{assumption}[Population support overlap]
 \label{assump4}
$0< \pi(X) <1$. 
\end{assumption}

Under  assumptions \ref{assump0}--\ref{assump4}, the treatment effect in the external population $\xi$ is 
identified by
 $$\xi=E\{E(Y\mid X,D=1,T=1)-E(Y\mid X,D=1,T=0)\mid D=0\},$$
and the overall treatment effect $\psi$ is identified by $\psi=\tau \cdot q+\xi \cdot  (1-q)$. 
For identification of $\psi$  and $\xi$,   it would be sufficient to assume  the mean exchangeability for  $Y_1-Y_0$, under which the efficient estimators of  $\psi$  and $\xi$ are  the one proposed by \citet{dahabreh2019generalizing} and the one by \cite{rudolph2017robust,dahabreh2020extending}, respectively. 
However, as we show later they can not make use of the outcome information of external samples and thereby do not admit efficiency improvement with full external data. 
Under the nonparametric model where the observed data distribution is restricted only by assumptions \ref{assump0}--\ref{assump4}, we derive the the efficient influence functions  for $\xi$ and $\psi$, respectively.

\begin{proposition}\label{oefficiency}
Under assumptions \ref{assump0}--\ref{assump4}, the efficient influence functions  for $\psi$ and  $\xi$ are 
\begin{equation*}
\begin{aligned}
\mathrm{IF}_\psi  &= \Delta(X)-\psi
+\frac{1}{\pi(X)}\left\{\dfrac{DT}{p(X)} R_1-  W(X,D,T)R_0\right\},\\
\mathrm{IF}_\xi &= \frac{1}{1-q}\left[(1-D)\Big\{\Delta(X)-\xi\Big\}
+\frac{1-\pi(X)}{\pi(X)}\left\{\dfrac{DT}{p(X)} R_1-W(X,D,T)R_0\right\}\right],
\end{aligned}
\end{equation*}
respectively, and the semiparametric efficiency bounds for estimating  $\psi $ and $\xi$  are $B_\psi =E(\mathrm{IF}_\psi^2)$  and  $B_\xi =E(\mathrm{IF}_\xi^2)$, respectively. 
\end{proposition}

The efficient influence functions $\mathrm{IF}_\psi$ and $\mathrm{IF}_\xi$ motivate the following  estimators, 
\begin{equation*}
\begin{split}
\hat\psi_{\rm dr}&=\hat{E}\left\{
\hat\Delta(X) + \dfrac{DT}{{\pi}(X;\widehat{\alpha})p(X;\hat{\phi})}\widehat R_1
-\dfrac{\widehat W(X,D,T)}{{\pi}(X;\hat{\alpha})}\widehat R_0\right\},\\
\hat\xi_{\rm dr}&= \frac{1}{1-\hat q}\hat{E}\left[(1-D) \widehat\Delta(X)+
\frac{1-\pi(X;\hat{\alpha})}{\pi(X;\hat{\alpha})} \left\{\dfrac{DT}{p(X;\hat{\phi})}\widehat R_1
-{\widehat W(X,D,T)}\widehat R_0\right\}\right] .
\end{split}
\end{equation*}
Semiparametric estimators $\hat\psi_{\rm dr}$ and $\hat\xi_{\rm dr}$ are both locally efficient, in the sense that the corresponding asymptotic variance attains the efficiency bound when all working models are correctly specified.
In addition, they  also enjoy the doubly robustness property.

\begin{proposition}\label{p2}
Under assumptions~\ref{assump0}--\ref{assump4} and regularity conditions described in theorems 2.6 and 3.4 of \citet{newey1994large}, the
 estimators $\hat \psi_{\mathrm{dr}}$ and   $\hat \xi_{\mathrm{dr}}$ are  consistent and asymptotically normal if either 
(i) the outcome models  $m_t(X;\beta_t)$ for $t \in \{0,1\}$ are correct,  or 
(ii) the propensity score models $\pi(X; \alpha)$ and $p(X; \phi)$ are correct.
\end{proposition}

Analogous to the estimator $\hat\tau_{\rm dr}$, consistency of $\hat \psi_{\mathrm{dr}}$ or $\hat \xi_{\mathrm{dr}}$  does not depend on correct specification of $r(X)$.
If  $r(X)$ is replaced by zero and $\beta_0$ is estimated with only the trial data,  $\hat\psi_{\rm dr}$  and $\hat\xi_{\rm dr}$  reduce to the estimator
proposed by \citet{dahabreh2019generalizing} and the one by \cite{rudolph2017robust,dahabreh2020extending}, hereafter referred to as $\tilde\psi_{\rm dr}$ and $\tilde\xi_{\rm dr}$, respectively.
However, $\tilde\psi_{\rm dr}$ and $\tilde\xi_{\rm dr}$ only utilize   the baseline covariates available  in the external data,
whereas our  estimators  can incorporate the information of both the  outcome and covariates in the external data.
Therefore,  $\hat \psi_{\mathrm{dr}}$ and $\hat \xi_{\mathrm{dr}}$ are more efficient. 
When  working models are correctly specified,  the   asymptotic variance of $\hat\psi_{\rm dr}$  is 
smaller than that of  $\tilde\psi_{\rm dr}$  by
\begin{equation*} 
E\left[ \left\{  \frac{1}{\pi(X)\{1-p(X)\}} - \frac{1}{\pi(X)\{1-p(X)\}+\{1-\pi(X)\} r(X)}  \right\} V_1(X)\right],
\end{equation*}
and asymptotic variance of $\hat\xi_{\rm dr}$  is 
smaller than that of   $\tilde\xi_{\rm dr}$   by
\begin{equation*} 
E\left[ \left\{ \dfrac{\{1-\pi(X)\}^2}{\pi(X)\{1-p(X)\}} - \dfrac{\{1-\pi(X)\}^2}{\pi(X)\{1-p(X)\}+\{1-\pi(X)\}r(X)} \right\} \frac{V_1(X)}{(1-q)^2}\right].
\end{equation*}

\section{Simulation study} \label{simu}

We evaluate the performance of the proposed methods via simulations. 
We consider four  data generating scenarios  with a continuous response:
(i) both the set of outcome models  and the set of propensity score models are correct, (ii) only the set of outcome models  are correct, (iii) only the set of propensity score models  are correct, and (iv) both    are incorrect.
In our settings, $\mbox{pr}(D=1)$ and $\mbox{pr}(T=1\mid D=1)$ are around 50\%. 
Details of the data generating mechanisms and the specification of  working models are  described in the Supporting Information.  

We use the following methods to estimate the average treatment effect $\tau$:  the full-data doubly robust estimator  $\hat{\tau}_{\mathrm{dr}}$ with a correct variance ratio, and the trial-based doubly robust estimator $\tilde{\tau}_{\mathrm{dr}}$. Besides, to investigate the impact of misspecification of the variance ratio, we also report  a full-data doubly robust estimator with a constant variance ratio.

\begin{figure}[H]
\hspace{1.6em}\includegraphics[scale=0.45]{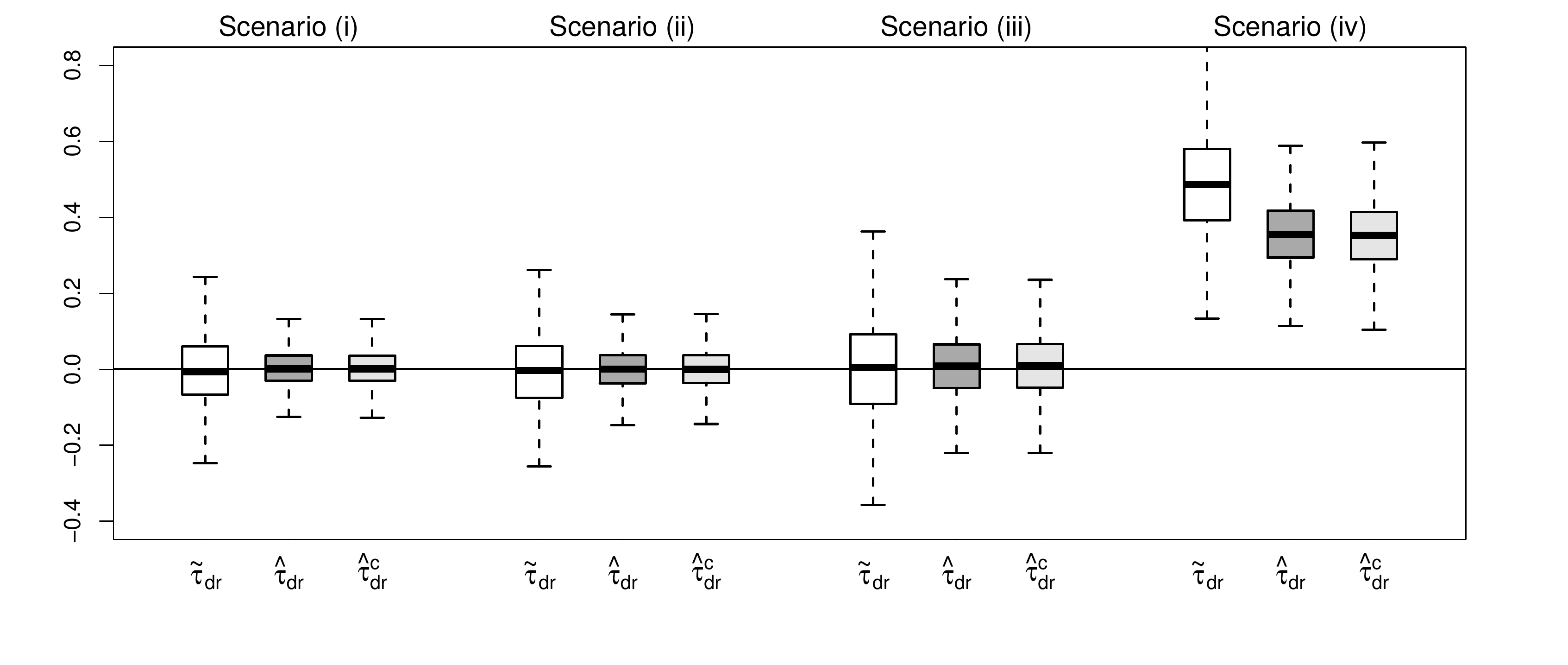}
\vspace{-1.6em}
\caption{Bias boxplots for estimators of the treatment effect $\tau$ in the clinical trial.  }\label{fig1}
\end{figure}

We simulate 1000 replicates under 1000 sample size for each scenario and summarize the results with bias boxplots in Fig.~\ref{fig1}.  
For  estimation  of $\tau$, the proposed full-data estimators have smaller variability, which indicates that leveraging information from the external data improves the efficiency. 
As expected, all three estimators have very small bias if either the set of outcome models or the set of  propensity score models  are correct.
But when both sets of working models are incorrect,  these estimators  may be biased; in this case,   the proposed full-data estimators have smaller bias than the trial-based estimator   in our simulations, 
although, this may not  hold in general.
Moreover,  although the variance ratio is not a constant in the data generating process,  the performance of the full-data doubly robust estimator  is not compromised by using  a constant variance ratio.
Table \ref{cvp} shows coverage probabilities of the $95\%$ confidence interval based on normal approximation. 
The coverage probability approximates the nominal level of $0.95$ provided that either of the working models is correct. 
We also provide simulation results for estimation of   $\psi$ and $\xi$ in the 
Supporting Information, showing that the full-data estimators $\hat{\psi}_{\mathrm{dr}}$
and $\hat{\xi}_{\mathrm{dr}}$ have very small bias even if one of the working models is incorrect, and have smaller variability than $\tilde{\psi}_{\mathrm{dr}}$
and $\tilde{\xi}_{\mathrm{dr}}$, respectively. 

In the Supporting Information, we  conduct additional numerical simulations  particularly under the scenarios where the mean exchangeability is violated. We fit a parametric outcome model for $E(Y\mid D,X,T=0)$ and then test interactive effects between $D$ and $X$ to assess the mean exchangeability.
The simulation results show that in certain cases where the engagement effects are weak, the full-data estimator $\hat{\tau}_{\mathrm{dr}}$ may still have  a smaller MSE than the trial-based estimator  $\tilde{\tau}_{\mathrm{dr}}$. 
In the presence of  strong engagement effects, the bias is no longer negligible, however, can be detected via the proposed test of the mean exchangeability.

\begin{table}
	\centering
	\def\~{\hphantom{0}}
	  	 \begin{minipage}{95mm}
	\caption{ \label{cvp} Coverage probability of $95\%$ confidence interval} 
	%\fbox{%
	\begin{tabular}{@{}lp{3em}p{20mm}p{20mm}p{20mm}}
	   Scenario& & & $\tau$ &  \\ 
	   && $\tilde{\tau}_{\mathrm{dr}}$  & $\hat{\tau}_{\mathrm{dr}}$  & $\hat{\tau}_{\mathrm{dr}}^c$    \\
	  (i) &&{0.955}  & {0.954} & {0.959}  \\ 
	  (ii) && {0.946}  & {0.955} & {0.956}      \\
	   (iii) && {0.960} & {0.956} & {0.952}  \\
	  (iv) && {0.096} & {0.031} & {0.046}     \\
	  \end{tabular} 
	  \vspace{2em}
	  \begin{tablenotes}
	  	\small
	  	\item Note for Fig.~\ref{fig1} and table \ref{cvp}:  $\hat{\tau}_{\mathrm{dr}}^c$ is  obtained by using using a constant variance ratio  in the  full-data doubly robust estimation of $\tau$.
	  	\end{tablenotes}
	  \end{minipage}   
\end{table}

  \section{An H.pylori infection example} \label{ana}
  
  We apply our methods to the H.pylori infection dataset for illustration.
  The trial dataset  is obtained from a two-arm clinical trial conducted at the TCM hospital;  
  it consists of 362 observations, of which 180 patients are assigned to the triple therapy,  and the rest are assigned to a combination treatment including both the triple therapy and TCM.
  The external dataset is obtained from  a single-arm clinical trial conducted at the Western-style hospital; it contains  110 observations, of which all patients are assigned to the triple therapy. To ensure the  uniformity of treatment, for at least 2 weeks prior to treatment, patients in both trials are instructed not to take any other medications that may interact with a study drug or affect the clinical result, such as PPI, H2RA and bismuth.
  The treatment process lasted for four weeks, and the outcome of interest  is the binary disease status after the treatment detected by the C-14 urea breath test (UBT).  The baseline covariates are the same in both external and trial data, including age, gender, height, BMI, work type, education level, marriage status and information on patients' symptoms etc.; we provide   descriptive statistics and more descriptions in the Supporting Information.

  We are interested in whether the additional Chinese medicine  would improve the cure rate in the clinical trial, particularly, testing the null hypothesis  $H_0:$ $\tau\leq0$ against  $H_1:$ $\tau>0$. 
  All patients assigned to the triple therapy  receive identical treatment and dose, 
  and we assume the mean exchangeability for $Y_0$.
  We estimate  $\tau$  with the full-data estimator $\hat{\tau}_{\mathrm{dr}}$ and the trial-based estimator $\tilde{\tau}_{\mathrm{dr}}$. 
  The conditional outcome means and propensity scores are fitted with  logistic models.  We assess  the mean exchangeability by testing whether there exist interactive effects between $D$ and $X$ on $Y$ with the  logistic outcome model built  upon control samples. The $p$-value of the test is 0.441, and hence under the significance level of 0.05, we can not reject the assumption \ref{assump2}.
  Because the outcome is binary, the variance ratio $r(X)$ is set to  one in our analysis. 
  We also estimate the overall treatment effect $\psi$ with  $\hat\psi_{\rm dr}$ and $\tilde{\psi}_{\mathrm{dr}}$, and the treatment effect of the external data $\xi$ with  $\hat\xi_{\rm dr}$ and $\tilde{\xi}_{\mathrm{dr}}$.  Table \ref{tab2} presents the data analysis results, including point estimates, variance estimates, and $p$-values.

  For estimation of the treatment effect in the  trial population $\tau$, 
  the point estimates  $\hat{\tau}_{\mathrm{dr}}$ and $\tilde{\tau}_{\mathrm{dr}}$ are close;
  however, the variance of the full-data doubly roust estimator $(\hat{\tau}_{\mathrm{dr}})$ is only $82\%$ of that of 
  the trial-based estimator $(\tilde{\tau}_{\mathrm{dr}})$.  
  As a result,   $\hat{\tau}_{\mathrm{dr}}$ leads to a smaller $p$-value (0.073) than $\tilde{\tau}_{\mathrm{dr}}$ ($p$-value: 0.110). 
  For estimation of the treatment effects $\psi$,  
  the proposed estimator  $\hat\psi_{\rm dr}$ has a smaller variance and also a smaller $p$-value (0.041) than $\tilde{\psi}_{\mathrm{dr}}$ ($p$-value: 0.069). The claim also applies to the estimation of $\xi$.
  Therefore, under the significance level of 0.1, we reject the null hypothesis based on $\hat{\tau}_{\mathrm{dr}}$,  $\hat\xi_{\rm dr}$ and $\hat\psi_{\rm dr}$,  and conclude that the combination treatment may have a better efficacy than the standard triple therapy  in the trial population, the external population  and the 
  overall population.
  
   To assess the robustness of the findings, we incorporate only the top half of the external data into the analysis.  The resulting estimator $\hat{\tau}_{\rm dr}$ is  0.052 with variance $17.89\times 10^{-4}$; as expected, the  variance is smaller than that of the trial-based estimator ($19.55\times 10^{-4}$), but  larger  than that of the full-data estimator ($16.10\times 10^{-4}$).

\begin{table}
	\centering
	 \def\~{\hphantom{0}}
	 \begin{minipage}{175mm}
	\caption{\label{tab2} Estimates of the causal effects in H.pylori application } 
	%\fbox{%
\begin{tabular}{lccclccc}
			&point  est. {\footnotesize$\times 10^2$}  & variance {\footnotesize$\times 10^4$}  & $p$-value & &point est. {\footnotesize$\times 10^2$} & variance {\footnotesize$\times 10^4$} &  $p$-value  \\ 
			$\hat{\tau}_{\mathrm{dr}}$ & 5.82 & 16.10 & 0.073  &
			$\tilde{\tau}_{\mathrm{dr}}$& 5.43 & 19.55 & 0.110  \\
			$\hat\psi_{\rm dr}$ & 6.72 & 14.98 & 0.041 &
			$\tilde{\psi}_{\mathrm{dr}}$& 6.55 & 19.56 & 0.069  \\
			$\hat\xi_{\rm dr}$ & 9.67 & 21.50 & 0.019 &
			$\tilde{\xi}_{\mathrm{dr}}$& 10.24 & 38.05 & 0.048  \\
	\end{tabular}
	\end{minipage}
\end{table}

  \section{Discussion} \label{dis}
  
  Our estimators are not restricted to the study design of the H.pylori application. 
  The methods remain valid as long as the study design  fulfills the requirement of the data structure and assumptions.
  However, under different study designs, the trial and external populations may differ, and so may the treatment effects.
  Therefore, the researchers need to   first clarify  the population of interest upon which the inference is drawn. In the Supporting Information, we discuss the practical  implications of the effects of interest when the trial and external datasets are collected under different study designs.

  The use of external data may yield bias if the mean exchangeability is violated when there exist engagement effects.  
   We denote the selection bias by
   \begin{equation*}
   E(Y_0\mid X,D=1)-E(Y_0 \mid X,D=0)=b(X),
   \end{equation*}
   which encodes  the strength of  engagement effects within each level of $X$.
  Assuming sufficiently flexible working models are employed such that no approximation error is introduced by model misspecification,  we can show that the asymptotic bias of $\hat{\tau}_{\rm{dr}}$ is 
   \begin{equation*}
   \begin{aligned}
   \Lambda=  E\left[ \frac{\pi(X)}{q} \cdot  \frac{ \{1-\pi(X)\} r(X)}{\pi(X)\{ 1-p(X)\}+\{1-\pi(X)\}r(X)} \cdot b(X) \right].
   \end{aligned}
   \end{equation*}
Suppose $b(X)$ is bounded with $|b(X)|\leq B$, then we have $|\Lambda| \leqslant B$, which  states that the asymptotic bias  of $\hat{\tau}_{\rm{dr}}$ does  not exceed that the largest difference between the conditional means of two datasets.  
As a result,   weak engagement effects would not negate our inference, although large ones can; see the Supporting Information for more discussions  on this issue.

\section*{Acknowledgement}
This work is partially supported by Beijing Natural Science Foundation (Z190001), National Natural Science Foundation of China (12071015, 12026606 and 81773546) and National New Drug Innovation Program (2017ZX09304003). The authors greatly appreciate the helpful comments and suggestions from the editor,
the associate editor and two anonymous referees.

\section*{Supporting Information}
Web Appendices, Tables, and Figures referenced in Sections 2--6 are available with this paper at the Biometrics website on Wiley Online Library.  
 R codes to replicate the simulation results are also provided.
 
%  This section is optional.  Here is where you will want to cite
%  grants, people who helped with the paper, etc.  But keep it short!

%  Not included in the original version of this paper!

%  Here, we create the bibliographic entries manually, following the
%  journal style.  If you use this method or use natbib, PLEASE PAY
%  CAREFUL ATTENTION TO THE BIBLIOGRAPHIC STYLE IN A RECENT ISSUE OF
%  THE JOURNAL AND FOLLOW IT!  Failure to follow stylistic conventions
%  just lengthens the time spend copyediting your paper and hence its
%  position in the publication queue should it be accepted.

%  We greatly prefer that you incorporate the references for your
%  article into the body of the article as we have done here 
%  (you can use natbib or not as you choose) than use BiBTeX,
%  so that your article is self-contained in one file.
%  If you do use BiBTeX, please use the .bst file that comes with 
%  the distribution.

\bibliographystyle{biometrika.bst} 
\bibliography{mybibilo.bib}

%%%%%%%%%%%%%%%%%%%%%%%%%%%%%%%%%%%%%%%%%%%%%%%%%%%%%%%%%%%%
\end{document}